\documentclass[useAMS,usenatbib]{mn2e}
\usepackage{amsmath}
\usepackage{amssymb}
\usepackage{color}
\usepackage{times}
\usepackage{graphicx}
\usepackage{hyperref}
\newcommand{\apj}{Astrophys. J.}

\newcommand{\apjl}{Astrophys. J.}
\newcommand{\aap}{Astron. Astrophys.}
\newcommand{\icarus}{Icarus}
\newcommand{\mnras}{Mon. Not. R. Astron. Soc.}
\newcommand{\araa}{Ann. Rev. Astron. Astrophys.}
\newcommand{\apjs}{Astrophys. J. Supp.}

\title[Formation of Super-Earths and Mini-Neptunes]{The Formation of Super-Earths and Mini-Neptunes with Giant Impacts}
\author[N.K. Inamdar and H.E. Schlichting]{Niraj K. Inamdar\thanks{E-mail:
inamdar@mit.edu} and Hilke E. Schlichting\\
Department of Earth, Atmospheric and Planetary Sciences, Massachusetts Institute of Technology, Cambridge, MA 02139, USA}
\begin{document}

\date{Accepted 2015 January 6. Received 2014 December 14; in original form 2014 October 12}

\pagerange{\pageref{firstpage}--\pageref{lastpage}} \pubyear{2015}

\maketitle

\label{firstpage}

\begin{abstract}
The majority of discovered exoplanetary systems harbour a new class of planets, bodies that are typically several times more massive than the Earth but that orbit their host stars well inside the orbit of Mercury. The origin of these close-in super-Earths and mini-Neptunes is one of the major unanswered questions in planet formation. Unlike the Earth, whose atmosphere contains less than $10^{-6}$ of its total mass, a large fraction of close-in planets have significant gaseous envelopes, containing $1-10$ per cent or more of their total mass. It has been proposed that close-in super-Earths and mini-Neptunes formed \textit{in situ} either by delivery of $50-100M_{\oplus}$ of rocky material to the inner regions of the protoplanetary disc, or in a disc enhanced relative to the minimum mass solar nebula. In both cases, the final assembly of the planets occurs via giant impacts. Here we test the viability of these scenarios. We show that atmospheres that can be accreted by isolation masses are small (typically $10^{-3}-10^{-2}$ of the core mass) and that the atmospheric mass-loss during giant impacts is significant, resulting in typical post-giant impact atmospheres that are $8 \times 10^{-4}$ of the core mass. Such values are consistent with terrestrial planet atmospheres but more than an order of magnitude below atmospheric masses of $1-10$ per cent inferred for many close-in exoplanets. In the most optimistic scenario in which there is no core luminosity from giant impacts and/or planetesimal accretion, we find that post-giant impact envelope accretion from a depleted gas disc can yield atmospheric masses that are several per cent the core mass. If the gravitational potential energy resulting from the last mass doubling of the planet by giant impacts is released over the disc dissipation time-scale as core luminosity, then the accreted envelope masses are reduced by about an order of magnitude. Finally we show that, even in the absence of type I migration, radial drift time-scales due to gas drag for many isolation masses are shorter than typical disc lifetimes for standard gas-to-dust ratios. Given these challenges, we conclude that most of the observed close-in planets with envelopes larger than several per cent of their total mass likely formed at larger separations from their host stars.
\end{abstract}


\section{Introduction}
In recent years, it has become clear that planets more massive than Earth but less massive than Neptune are amongst the most common planets in our Galaxy \citep{HowardMarcyJohnson,Borucki2011,Batalha2013}. Based on their bulk density, it has been inferred that many of these planets, which typically orbit their host stars at semimajor axes of less than $0.1~\mathrm{AU}$, consist of rocky or icy cores surrounded by large H/He envelopes \citep{AdamsSeager,LopezFortney}. In some cases, these envelopes comprise tens of per cent or more of the overall planet mass and have radii commensurate with the core radius (by comparison, Earth's atmosphere makes up about $10^{-6}$ of its total mass and $10^{-3}$ its total radius). How these close-in planets formed, however, remains an outstanding question in planet formation. 

In order to understand the origin of super-Earths and mini-Neptunes, recent work has considered the \textit{in situ} accretion of gas envelopes by fully-formed cores of several Earth masses. \citet{LopezFortneyMiller} focused on modelling the thermal evolution of mini-Neptunes to determine the composition and structure of the Kepler-11 system. \citet{LopezFortney} investigated the thermal evolution and structure of mini-Neptunes in general, and calculated mass-radius relationships that were then used to determine core and envelope masses and radii for a number of observed exoplanets. \citet{BodenheimerLissauer} and \citet{Rogers2011} considered the accretion and evolution of mini-Neptunes over a variety of semimajor axes. In their models, the core luminosity is provided by accreting planetesimals (which do not appreciably change the core mass), and they conclude that formation further out in the disc is likely. \citet{IkomaHori} likewise considered gas accretion by a fully-formed core in the presence of a dissipating gas disc, assuming that the planetary cores first migrated inwards and accreted their envelopes \textit{in situ}. 

A common assumption of all these investigations is that the planetary cores have already formed before they proceed to accrete their gaseous envelopes. However, planet formation is likely to have passed through several, distinct phases [e.g. \citep{Safronov,Goldreich2004}], and the core masses that can be attained in the inner disc by accreting all the solids locally available are typically much less than that of the Earth. 
Assuming a shear-dominated velocity dispersion, the isolation mass $M_{\mathrm{iso}}$ that can form in the absence 
of migration at a semimajor axis $a$ is given by 
\begin{equation}\label{eq:IsoMassEq}
M_{\mathrm{iso}} = \left[\frac{4\pi a^2 C \Sigma_s}
								{\left(3M_{\star}\right)^{1/3}}\right]^{3/2},
\end{equation}
where $M_{\star}$ is the mass of the host star and $\Sigma_s$ is the surface density of solids in the disc. $C$ is a constant that describes the radial extent of the isolation mass's feeding zone in units of Hill radius. The Hill radius, which defines the extent over which the gravitational force of the planet overcomes tidal interactions with the host star, is given by $r_H = a \left(M_{\mathrm{iso}}/3M_{\star}\right)^{1/3}$.

In the outer disc, at several tens of AU, $M_{\mathrm{iso}}$ is on the order of a Neptune mass for a minimum mass solar nebula (MMSN) type disc \citep{Hayashi}. In the inner disc, however, the corresponding isolation masses are only a fraction of an Earth mass. Forming terrestrial planets or larger cores in the inner disc therefore requires an additional stage of assembly, in which collisions called giant impacts \citep{ChambersWetherill1998,Agnor1999} between protoplanets ultimately result in the formation of several terrestrial planet-mass bodies \citep{Chambers2001,Raymond2004}.
Furthermore, while it may seem possible that an arbitrary increase in disc solid density can allow for isolation masses in the inner disc of several $M_{\oplus}$, it has been shown that for standard gas-to-dust ratios, the corresponding gas disc is unstable for a significant fraction of observed exoplanets within $0.1~\mathrm{AU}$ of their host star \citep{Schlichting2014}. 

The necessity of giant impacts to assemble sufficiently large masses in the inner disc has motivated the development of models in which close-in super-Earths and mini-Neptunes form \textit{in situ} via giant impacts. In one proposed scenario, $50-100M_{\oplus}$ of rocky material is delivered to the inner part of the protoplanetary disc \citep{HansenMurray}, while in another, the protoplanetary disc has surface densities enhanced relative to the MMSN \citep{ChiangLaughlin}. In both cases, the final assembly of the planets occurs via giant impacts. Here we test the viability of such \textit{in situ} formation scenarios by examining how much gaseous envelope could have been accreted by planetary embryos before giant impacts, how much could be retained throughout the giant impact phase, and how much could have subsequently been accreted from a depleted disc after the giant impact phase. 

This paper is structured as follows. In Section \ref{sec:IsoAccrete}, we calculate atmospheric masses accreted by isolation masses for various semimajor axes. In Section \ref{sec:HydroEscape}, we calculate both the local and global atmospheric mass-loss as a function of impactor mass and velocity for planets with various atmosphere-to-core mass ratios. In Section \ref{sec:CollModel}, we construct a giant impact history and calculate atmospheric masses that can be retained throughout the giant impact phase using the results from Section \ref{sec:HydroEscape}. In Section \ref{sec:PostGiant}, we explore the possibility that envelope accretion occurred after giant impacts in a depleted gas disc. Discussion and conclusions follow in Section \ref{sec:DiscConc}.

\section{Accretion of Envelopes by Isolation Masses}\label{sec:IsoAccrete}
\subsection{Isolation Masses}\label{sec:IsoMasses}
To determine $M_{\mathrm{iso}}$ for a typical close-in planet, we first calculate the minimum surface density of solids $\Sigma_s$ in the disc needed to form the planet \textit{in situ} with giant impacts. For a planet of mass $M_p$, $\Sigma_s = M_p/\left( 2\pi a \Delta a\right)$, where for giant impacts, the annulus width $\Delta a \simeq 2 v_{\mathrm{esc}}/\Omega$  \citep{Schlichting2014}. Here, $v_{\mathrm{esc}} = \sqrt{2GM_p/R_p}$ is the escape velocity of the planet and $\Omega = \sqrt{GM_{\star}/a^3}$ is the Keplerian frequency of its orbit.
To calculate $v_{\mathrm{esc}}$, we use the mass-radius relationship \citep{LissauerArch}
\begin{equation}\label{eq:MassRadiusNep}
M_p = M_{\oplus}\left(\frac{R_p}{R_{\oplus}}\right)^{2.06},
\end{equation}
which is appropriate for Earth- to Neptune-sized planets. Using Eq. \eqref{eq:IsoMassEq}, we then calculate the corresponding isolation mass $M_{\mathrm{iso}}$.

\begin{figure}
	\centering
	\includegraphics[width=0.5\textwidth]{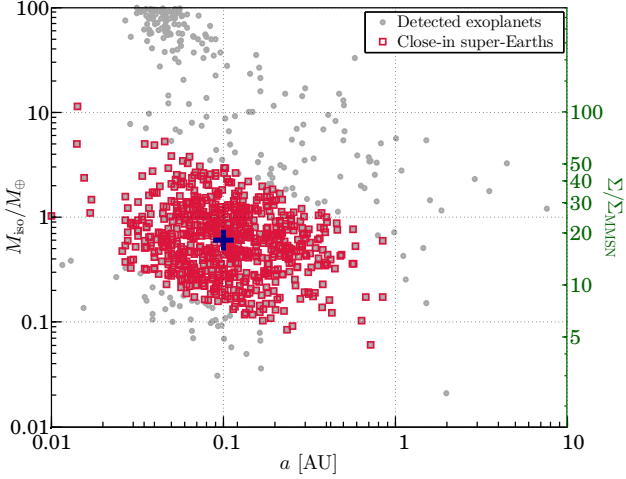}
	\caption{ 
Isolation masses $M_{\mathrm{iso}}$ for detected exoplanets at their current orbital distances. Isolation masses $M_{\mathrm{iso}}$ for all detected exoplanets relative to an Earth mass, $M_{\oplus}$, are shown as grey circles. Close-in super-Earths and mini-Neptunes, exoplanets with semimajor axes less than 1 AU and with masses greater than an Earth mass but less than a Neptune mass, are highlighted as red squares. On the right-hand axis, we map $\Sigma_s$ for isolation masses to a protoplanetary disc surface density enhancement relative to that of the minimum mass solar nebula (denoted $\Sigma/\Sigma_{\mathrm{MMSN}}$). For a typical close-in planet with an assembled mass of about $4.5M_{\oplus}$ at 0.1 AU, $M_{\mathrm{iso}}$ has a median value of $0.6 M_{\oplus}$ (blue cross). Exoplanet data taken from \url{exoplanet.eu} data base \citep{ExoEuCat}.}
	\label{fig:Fig1}
\end{figure}

In Fig. \ref{fig:Fig1}, we show the isolation masses for observed exoplanets. Assuming that the planets formed at their observed semimajor axis and that their isolation masses were later merged by giant impacts to yield the observed \textit{Kepler} planets, the isolation masses for close-in planets are between $0.1$ and $5M_{\oplus}$. We also map the corresponding disc surface density $\Sigma_s$ to an enhancement relative to that of the MMSN, which we denote $\Sigma/\Sigma_{\mathrm{MMSN}}$. For a typical close-in planet with an assembled mass of about $4.5M_{\oplus}$ at 0.1 AU, $M_{\mathrm{iso}}$ has a median value of $0.6 M_{\oplus}$, corresponding to a disc enhanced relative to the MMSN by a factor of roughly $\Sigma/\Sigma_{\mathrm{MMSN}} = 20$ (blue cross). 

\subsection{Atmospheric Accretion Model}\label{sec:ModelDetails}

The envelope mass $M_{\mathrm{atm}}$ is given by
\begin{equation}
M_{\mathrm{atm}} = 4\pi\int_{R_{\mathrm{core}}}^{r_{\mathrm{min}}}\rho(r) r^2 dr,
\end{equation}
where $r$ is the distance measured outward from the centre of the core, $R_{\mathrm{core}}$ is the radius of the core, and $\rho$ is the density profile of the envelope. 
The amount of gas that an isolation mass core can accrete also depends on the extent of its Hill or Bondi radius, the smaller of which is denoted $r_{\mathrm{min}}$. The Bondi radius $r_B$ describes the range over which the escape velocity of the core is greater than the sound speed $c_s$ of the gas. If $G$ is the gravitational constant, then $r_B = 2GM_{\mathrm{core}}/c_s^2$. The core mass $M_{\mathrm{core}}^{*}$ at which $r_{\mathrm{min}}$ transitions from the Bondi radius to the Hill radius is calculated by setting $r_B = r_H$, yielding 
\begin{equation}\label{eq:M_trans}
M_{\mathrm{core}}^{*} = \left[\frac{a \gamma_a k_B T_{\mathrm{disc}}}
	{2G \mu m_p \left(3M_{\star}\right)^{1/3}}\right]^{3/2}.
\end{equation}
Here, $k_B$ is the Boltzmann constant, $\mu$ is the molecular mass of the gas (taken to be 2.34), $m_p$ is the proton mass, and $\gamma_a$ is the adiabatic index of the gas, which we assume is $7/5$, appropriate for a diatomic gas. Eq. \eqref{eq:M_trans} implies that if $M_{\mathrm{core}}$ is less than $M_{\mathrm{core}}^{*}$, then $r_{\mathrm{min}}$ is given by $r_B$, and if $M_{\mathrm{core}}$ is greater than $M_{\mathrm{core}}^{*}$, then $r_{\mathrm{min}}$ is given by $r_H$. We assume that the core radius, $R_{\mathrm{core}}$, is related to $M_{\mathrm{core}}$ as 
\begin{equation}\label{eq:MassRadius} 
R_{\mathrm{core}} \approx R_{\oplus}\left(\frac{M_{\mathrm{core}}}
											{M_{\oplus}}\right)^{1/4},
\end{equation}
which is consistent with results from both observations \citep{LopezFortney,LissauerPacked} and planetary interior mass-radius models for planets composed primarily of rocky material \citep{SeagerMassRadius}.
Eq. \eqref{eq:MassRadius} also implies that for some $M_{\mathrm{core}}$, $r_{\mathrm{min}}$ becomes smaller than $R_{\mathrm{core}}$, in which case the isolation mass cannot accrete any envelope. In the inner disc, where $M_{\mathrm{core}} < M_{\mathrm{core}}^{*}$, this occurs when
\begin{equation}\label{eq:core_no_accrete}
M_{\mathrm{core}} = \left(\frac{R_{\oplus}c_s^2}{2GM_{\oplus}^{1/4}}\right)^{4/3}.
\end{equation}

We assume that the surface density of solids in the disc is given by $\Sigma_{s} = \Sigma_{s,1}\left(a/\mathrm{AU}\right)^{-3/2}$, while the surface density of gas is given by $\Sigma_g = \Sigma_{g,1}\left(a/\mathrm{AU}\right)^{-3/2}$. For accretion of gaseous envelopes, we assume throughout this paper that gas surface densities are enhanced by a factor of 200 relative to the solid disc surface densities so that for a disc enhanced relative to the MMSN by a factor of 20, $\Sigma_{s,1} = 140~\mathrm{g/cm^2}$ and $\Sigma_{g,1} = 28,000~\mathrm{g/cm^2}$. In our disc model, the temperature $T_{\mathrm{disc}}$ is assumed to be uniform at $1500~\mathrm{K}$ out to 0.1 AU, beyond which $T_{\mathrm{disc}}$ scales as $a^{-2/3}$ \citep{DAlessio}. The vertical structure of the disc is assumed locally isothermal, so that if $\Omega$ is the Keplerian angular velocity, the scale height $H = c_s/\Omega$ and $\rho_{\mathrm{disc}} \simeq \Sigma_g/\left(2H\right)$. When calculating atmospheric structure, care must be taken to ensure $r_{\mathrm{min}} < H$, since for $r_{\mathrm{min}} > H$, the core will no longer accrete spherically. For our disc model, we have verified that for semimajor axes of typical close-in Kepler planets ($0.03-0.3~\mathrm{AU}$; see, e.g., Fig. \ref{fig:Fig1}) $r_{\mathrm{min}}$ is less than $H$ for the vast majority of isolation masses shown in Fig. \ref{fig:Fig1}, such that our assumption of spherical accretion is valid. We find that at $0.03~\mathrm{AU}$, $r_{\mathrm{min}}$ becomes comparable to $H$ \textit{only} for isolation masses of $4M_{\oplus}$ and larger.

\subsubsection{Analytically-Calculated Adiabatic Envelope Masses}
To calculate the structure of the accreted envelope, we use the disc density, pressure, and temperature ($\rho_{\mathrm{disc}}$, $p_{\mathrm{disc}}$, and $T_{\mathrm{disc}}$, respectively) as boundary conditions at $r_{\mathrm{min}}$ and integrate the equations of hydrostatic equilibrium inward to $R_{\mathrm{core}}$. The accreted envelope will typically have an outer layer in radiative equilibrium with the disc, while the inner portion of the envelope is in adiabatic equilibrium. In the inner disc ($\lesssim 0.1~\mathrm{AU}$), where disc gas densities and opacities are sufficiently high, the radiative layer of the envelope contributes negligibly to the total atmospheric mass, and the envelope can be regarded as essentially adiabatic \citep{Rafikov2006}. In this case, the density, pressure, and temperature of the envelope are given by
\begin{subequations}\label{eq:adiabatic_eqs}
\begin{align}
\frac{\rho}{\rho_{\mathrm{disc}}} &= \left[\frac{\gamma_a - 1}{\gamma_a}\lambda\left(\frac{r_{\mathrm{min}}}{r} - 1\right) + 1\right]^{1/(\gamma_a - 1)}, \\
\frac{p}{p_{\mathrm{disc}}}       &= \left[\frac{\gamma_a - 1}{\gamma_a}\lambda\left(\frac{r_{\mathrm{min}}}{r} - 1\right) + 1\right]^{\gamma_a/(\gamma_a - 1)}, \\
\frac{T}{T_{\mathrm{disc}}}       &=     ~~\frac{\gamma_a - 1}{\gamma_a}\lambda\left(\frac{r_{\mathrm{min}}}{r} - 1\right) + 1,
\end{align}
\end{subequations}
where $r$ is measured outward from the core centre and where $\lambda = GM_{\mathrm{core}}\mu m_p/\left(r_{\mathrm{min}} k_B T_{\mathrm{disc}}\right)$ is the escape parameter at $r_{\mathrm{min}}$. For an adiabatic envelope, to good accuracy, we can approximate the envelope mass $M_{\mathrm{atm}}$ as simply $4\pi \rho_{\mathrm{disc}} r_{\mathrm{min}}^3/3$. When $M_{\mathrm{core}} \lesssim M_{\mathrm{core}}^*$ (so that $r_{\mathrm{min}} = r_B$), the envelope to core mass ratio can therefore be approximated by
\begin{eqnarray}\label{eq:BondiApprox}
\frac{M_{\mathrm{atm}}}{M_{\mathrm{core}}} \simeq 1.10\times 10^{-2} \left(\frac{\Sigma_{g,1}}{28,000~\mathrm{g/cm^2}}\right)\left(\frac{T_{\mathrm{disc}}}{1500~\mathrm{K}}\right)^{-7/2} \times \nonumber \\
 \hfill\left(\frac{a}{0.1~\mathrm{AU}}\right)^{-3} \left(\frac{M_{\mathrm{core}}}{M_{\oplus}}\right)^2.
\end{eqnarray}
On the other hand, if $M_{\mathrm{core}}\gtrsim M_{\mathrm{core}}^{*}$, so that $r_{\mathrm{min}} = r_{H}$, we have
\begin{equation}\label{eq:HillApprox}
\frac{M_{\mathrm{atm}}}{M_{\mathrm{core}}} \simeq 3.08\times 10^{-2} \left(\frac{\Sigma_{g,1}}{28,000~\mathrm{g/cm^2}}\right) \left(\frac{T_{\mathrm{disc}}}{1500~\mathrm{K}}\right)^{-1/2}.
\end{equation}
Fig. \ref{fig:Fig2} demonstrates the accuracy of Eqs. \eqref{eq:adiabatic_eqs} compared to the more rigorous calculations we detail below.

\subsubsection{Numerically-Calculated Radiative-Convective Envelope Masses}
Although in the inner disc the accreted envelope may be regarded as adiabatic, in the outer disc or in a partially depleted disc (see Section \ref{sec:PostGiant}), the radiative component of the atmosphere contributes significantly to the total atmospheric mass. When this is the case, we must numerically integrate the equations of hydrostatic equilibrium and flux conservation. For our numerical calculations, the density, pressure, and temperature profiles of accreted envelopes are calculated using a fourth-order Runge-Kutta method. If we assume an ideal gas equation of state, then the equation of hydrostatic equilibrium
\begin{equation}\label{eq:HydroStat}
\frac{dp}{dr} = -\frac{GM_{\mathrm{core}}}{r^2}\rho,
\end{equation}
and that of flux conservation
\begin{equation}\label{eq:ThermEquil}
\frac{dT}{dr} = -\frac{3\kappa_R \rho}{16\sigma T^3} \frac{L}{4\pi r^2},
\end{equation}
allow us to solve for the envelope structure. $\kappa_R$ is the Rosseland mean opacity of the gas in the envelope, $\sigma$ is the Stefan-Boltzmann constant, and $L$ is the luminosity in the envelope. Eq. \eqref{eq:ThermEquil} implies radiative equilibrium, which will not be the case if the lapse rate is such that the envelope is locally unstable to convection. Thus, during numerical integration, if for a given $r$ Eqs. \eqref{eq:HydroStat} and \eqref{eq:ThermEquil} imply $d\log T/d\log P \geq \left(\gamma_a - 1\right)/\gamma_a$, then we replace Eq. \eqref{eq:ThermEquil} with $d\log T/d\log P = \left(\gamma_a - 1\right)/\gamma_a$, so that locally, adiabatic equilibrium holds. Where this transition occurs in the envelope defines the radiative-convective boundary.

For the parts of the envelope in radiative equilibrium, we use the analytic expressions for dust grain opacity $\kappa_{R,\mathrm{dust}}$ and molecular scattering opacity $\kappa_{R,\mathrm{mol}}$ given by \citet{Zhu} to calculate $\kappa_R$. At temperatures $\gtrsim 1800~\mathrm{K}$, dust grains sublimate, and we expect molecular scattering to be the dominant source of opacity \citep{Movshovitz}. At temperatures lower than this, both molecular scattering and grain opacity are accounted for. We assume in the latter case that opacities can be summed, so that $\kappa_R\rho = \left(\kappa_{R,\mathrm{dust}}/200 + \kappa_{R,\mathrm{mol}}\right)\rho$. The factor of $1/200$ arises because we assume that the density of dust grains in the atmosphere is 200 times less than the gas, consistent with the dust to gas ratio in the disc. Grain opacity will tend to dominate in the cooler outer envelope, which is also the portion of the envelope in radiative equilibrium with the disc.

\subsubsection{Accretion Luminosities}\label{sec:accLum}
The luminosity $L$ can arise from several possible sources. Luminosity may be provided by the core (either through radiogenic heating or by energy released from core accretion) and/or from changes in the gravitational and internal energy of the envelope during cooling and contraction. We therefore consider two limiting cases for the luminosity.\\
~\\ 
\textbf{Case 1.} In the first case, we assume that the luminosity is due to core accretion. $L$ is given by $L_{\mathrm{acc}} = G M_{\mathrm{core}}\dot{M}_{\mathrm{core}}/R_{\mathrm{core}}$, where $\dot{M}_{\mathrm{core}}$ is the core mass accretion rate. We take $\dot{M}_{\mathrm{core}} = M_{\mathrm{core}}/\tau_{\mathrm{acc}}$, where, following Eq. (A2) of \citet{Rafikov2006}, the accretion time-scale $\tau_{\mathrm{acc}}$ appropriate for an isolation mass is given by $\tau_{\mathrm{acc}} =  30~\mathrm{kyr} \times \left(M_{\mathrm{core}}/M_{\oplus}\right)^{1/3} \left(a/\mathrm{0.1~AU}\right)^3$. Using Eq. \eqref{eq:MassRadius}, we find 
\begin{equation}\label{eq:Acctime-scale}
L_{\mathrm{acc}} = 3.94\times 10^{27}~\frac{\mathrm{erg}}{\mathrm{s}} \times \left(\frac{M_{\mathrm{core}}}{M_{\oplus}}\right)^{17/12} \left(\frac{a}{0.1~\mathrm{AU}}\right)^{-3}.
\end{equation}
For a typical close-in planet isolation mass of $0.6 M_{\oplus}$ at $0.1~\mathrm{AU}$, $L_{\mathrm{acc}} \approx 1.9 \times 10^{27}~\mathrm{erg/s}$.
We consider this case an upper limit on the accretion luminosity because the accretion time-scale is short relative to typical disc lifetimes. Using this limit essentially assumes that the start of planet formation is delayed and that accretion proceeds until the gas disc dissipates and giant impacts set in. When $L = L_{\mathrm{acc}}$, the cooling time-scale $\tau_{\mathrm{cool}}$ for the accreted envelope is on the order of $10^{7}-10^{8}$ yr \citep{Rafikov2006}, while the disc lifetime is on the order of $10^{6}$ yr [e.g., \citep{Hillenbrand}]. Hydrostatic equilibrium is established on the order of the sound crossing time $r_{\mathrm{min}}/c_s$ ($<1$ yr), so that in this case, we neglect the time evolution of the envelope structure and consider only hydrostatic equilibrium. 

To verify our methods, we compared our results to those reported by \citet{BodenheimerLissauer} and \citet{Rafikov2006}. Using the \citet{BodenheimerLissauer} accretion model, and the disc temperatures and surface densities used for their Runs 2, 1, and 0.5 (corresponding to accretion at semimajor axes of $2~\mathrm{AU}$, $1~\mathrm{AU}$, and $0.5~\mathrm{AU}$ by cores of mass $2.15M_{\oplus}$, $2.20M_{\oplus}$, and $2.20M_{\oplus}$, respectively), we find $M_{\mathrm{atm}}/M_{\mathrm{core}} = 0.031$, $0.010$, and $0.034$, respectively. \citet{BodenheimerLissauer} calculate $M_{\mathrm{atml}}/M_{\mathrm{core}} = 0.025$, $0.025$, and $0.017$, within a factor of $2-3$ of our results. It is likely that differences between the two models arise from different prescriptions of $r_{\mathrm{min}}$ and different opacity models. We also checked our calculations against those of \citet{Rafikov2006}. Using the disc, opacity, and accretion models described in \citet{Rafikov2006}, we find, for example, that for a $1M_{\oplus}$ core, $M_{\mathrm{atm}} = 1.02 \times 10^{26}~\mathrm{g}$ at $0.1~\mathrm{AU}$ assuming a slow accretion time-scale ($\tau_{\mathrm{acc}} = 30~\mathrm{kyr}$); $M_{\mathrm{atm}} = 7.17 \times 10^{24}~\mathrm{g}$ assuming a medium accretion time-scale at $1~\mathrm{AU}$ ($\tau_{\mathrm{acc}} = 140~\mathrm{kyr}$); and $M_{\mathrm{atm}} = 2.50 \times 10^{24}~\mathrm{g}$ assuming a fast accretion time-scale at $10~\mathrm{AU}$ ($\tau_{\mathrm{acc}} = 300~\mathrm{kyr}$). All three results are indistinguishable from those presented in Fig. 6 of \citet{Rafikov2006}. \\
\\
\textbf{Case 2.} In the second case, we assume $L_{\mathrm{acc}} = 0$. Here instead $L$ evolves according to
\begin{equation}\label{eq:entevo}
L = \int_{R_{\mathrm{core}}}^{\mathrm{RCB}}4\pi r^2 \rho \frac{dS}{dt}dr,
\end{equation}
where $\mathrm{RCB}$ denotes the radiative-convective boundary and $S$ is the specific entropy of the envelope at $r$ \citep{PisoYoudin}. In this case, $L$ is dominated by the change in the internal and gravitational energy of the envelope interior to the radiative-convective boundary as the envelope contracts. Following \citet{LeeChiangOrmel}, we assume that the envelope is initially adiabatic after which it cools and the radiative-convective boundary evolves inwards. We solve the equations of hydrostatic equilibrium for a range of $L$ and follow the time evolution of the envelope structure and mass using Eq. \eqref{eq:entevo}. It is necessary to cut off the envelope evolution and gas accretion at a time commensurate with the disc lifetime. For our pre-giant impact gas accretion calculations, we do this at $\tau_{\mathrm{disc}} \sim 2~\mathrm{Myr}$, though we note that our calculations are insensitive to this cutoff time to within a few $\mathrm{Myr}$.\\
\indent This case corresponds to an upper bound for the atmospheric mass a core can accrete. As the envelope cools, the radiative-convective boundary moves inwards and the envelope accretes more gas. The rate of cooling will be regulated by several factors: the depth of the radiative-convective boundary with respect to $r_{\mathrm{min}}$, the opacity of the radiative layer, and the atmospheric mass. The contribution of an additional luminosity from the core, however, effectively delays this cooling, limiting the additional atmospheric mass that can be added over the lifetime of the gas disc.

\subsection{Isolation Mass Gas Accretion Results}\label{sec:IsoMassResults}
\begin{figure}
	\centering
	\includegraphics[width=0.5\textwidth]{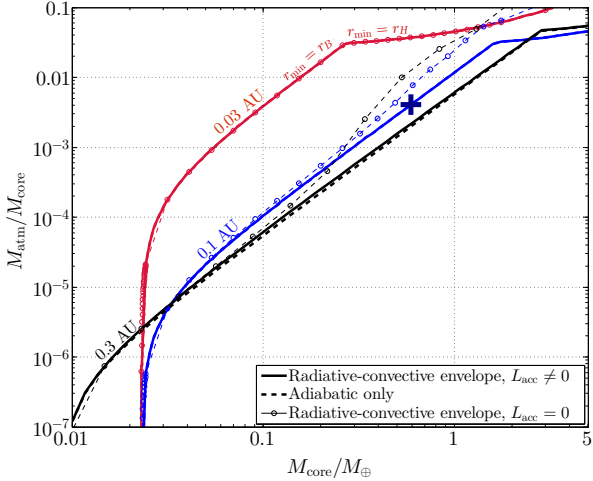}
	\caption{Atmospheric masses due to accretion of gas envelopes by isolation masses prior to the phase of giant impacts. Accreted envelope masses are shown for semimajor axes of 0.03 AU, 0.1 AU, and 0.3 AU. Atmospheric masses are calculated assuming a MMSN enhanced by a factor of 20 ($\Sigma_{g,1} = 28,000~\mathrm{g/cm^2}$, $\Sigma_{s,1} = 140~\mathrm{g/cm^2}$). 
In thick dashed lines, we show $M_{\mathrm{atm}}/M_{\mathrm{core}}$ for the case in which we assume that the atmosphere is fully adiabatic [i.e. using the analytic expressions Eqs. \eqref{eq:adiabatic_eqs}]. In thick solid lines, we show the results from numerical integrations assuming that the core accretion luminosity $L_{\mathrm{acc}}$ is nonzero and given by Eq. \eqref{eq:Acctime-scale}. In thin dashed lines with circular markers, we show $M_{\mathrm{atm}}/M_{\mathrm{core}}$ for the case in which $L_{\mathrm{acc}} = 0$ and $L$ is governed by Eq. \eqref{eq:entevo}. For an isolation mass and semimajor axis typical of close-in planets ($M_{\mathrm{core}} = 0.6M_{\oplus}$, $a = 0.1~\mathrm{AU}$, $\Sigma/\Sigma_{\mathrm{MMSN}} = 20$; see Fig. \ref{fig:Fig1}), $M_{\mathrm{atm}} \approx 4 \times 10^{-3} M_{\mathrm{core}}$ (blue cross). The difference in atmospheric mass between all the modelled cases is small. This is due to increased densities and hence opacities at smaller semimajor axes, which yields almost adiabatic envelopes and inhibits significant cooling for isolation mass cores.}
	\label{fig:Fig2}
\end{figure}

In Fig. \ref{fig:Fig2}, we show $M_{\mathrm{atm}}/M_{\mathrm{core}}$ calculated over a range of $M_{\mathrm{core}}$ consistent with close-in planet isolation masses, at semimajor axes of $0.03~\mathrm{AU}$, $0.1~\mathrm{AU}$, and $0.3~\mathrm{AU}$ (see, e.g. Fig. \ref{fig:Fig1}). In thick dashed lines, we show atmospheric masses calculated assuming a fully adiabatic envelope using Eqs. \eqref{eq:adiabatic_eqs}. In thick solid lines, we show atmospheric masses calculated numerically assuming $L = L_{\mathrm{acc}}$, where the accretion luminosity $L_{\mathrm{acc}}$ is given by Eq. \eqref{eq:Acctime-scale}. In thin dashed lines with circular markers, we show atmospheric masses calculated numerically assuming $L_{\mathrm{acc}} = 0$, with the evolution of $L$ governed instead by Eq. \eqref{eq:entevo}. For a typical close-in planet isolation mass of $M_{\mathrm{core}} = 0.6M_{\oplus}$ at $0.1~\mathrm{AU}$, we find $M_{\mathrm{atm}}/M_{\mathrm{core}} \approx 4 \times 10^{-3}$.

Fig. \ref{fig:Fig2} shows that in the inner disc envelopes are mostly adiabatic with only a small outer radiative layer that contains negligible mass. This is due to the increased gas densities and hence opacities in the inner disc. It is only at larger semimajor axes (e.g., the  $L_{\mathrm{acc}} = 0$ case at $0.3~\mathrm{AU}$ in Fig. \ref{fig:Fig2}) that we see a significant contribution to the total atmospheric mass from the outer radiative layer. Here, for cores that are sufficiently massive ($M_{\mathrm{core}}\gtrsim 0.3M_{\oplus}$), decreased opacities allow the envelope to cool, contract, and accrete larger atmospheric masses. 
 
For $a = 0.03~\mathrm{AU}$ and $0.1~\mathrm{AU}$, atmospheric masses drop off sharply at $M_{\mathrm{core}} = 0.023 M_{\oplus}$. Here, Eq. \eqref{eq:core_no_accrete} is satisfied, so that with our model parameters, no atmosphere can be accreted for $M_{\mathrm{core}} \lesssim 0.023 M_{\oplus}$. The break in the $0.01~\mathrm{AU}$ lines at about $M_{\mathrm{core}} = 0.25M_{\oplus}$, in the $0.1~\mathrm{AU}$ lines at $M_{\mathrm{core}} = 1.7M_{\oplus}$, and in the $0.3~\mathrm{AU}$ lines at $M_{\mathrm{core}} = 2.9M_{\oplus}$ indicate where $r_{\mathrm{min}}$ transitions from $r_B$ to $r_H$.

\section{Giant Impact-Induced Hydrodynamic Escape}\label{sec:HydroEscape}
\begin{figure}
	\centering
		\includegraphics[width=0.5\textwidth]{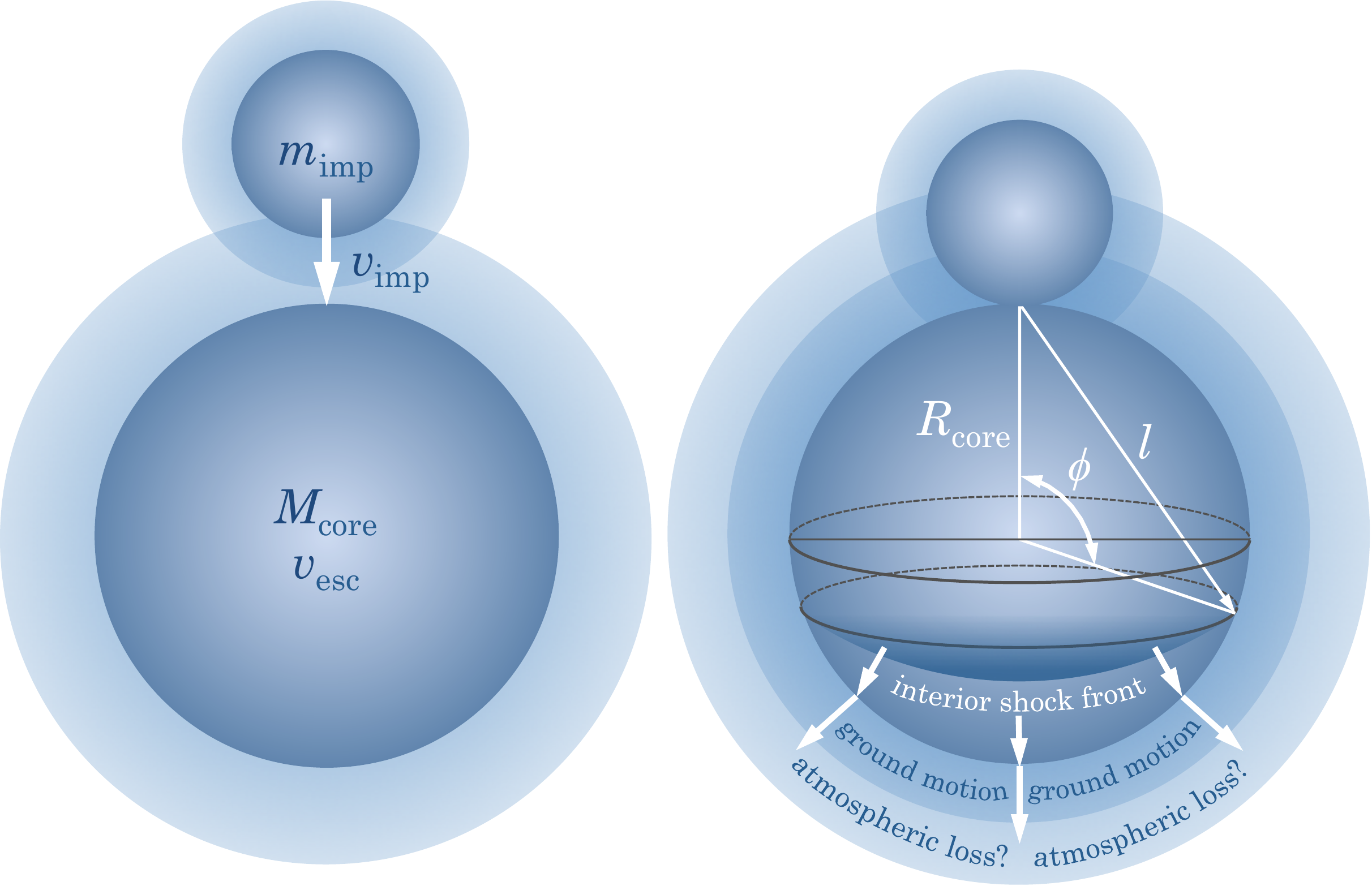}
	\caption{Schematic of giant impact-induced atmospheric loss. An impactor of mass $m_{\mathrm{imp}}$ and velocity $v_{\mathrm{imp}}$ approaches a planetary embryo of core mass $M_{\mathrm{core}}$ and escape velocity $v_{\mathrm{esc}}$ (left-hand panel). The collision generates a shock wave that 	propagates through the interior of the embryo core (right-hand panel). As this interior shock wave travels through the core, it subsequently initiates a global ground motion of the planet. This ground motion then launches a shock into the atmosphere above it, potentially leading to hydrodynamic escape.}
	\label{fig:Fig3}
\end{figure}

In Section \ref{sec:IsoAccrete}, we calculated the atmospheric structure and mass that an isolation mass accretes in the presence of the full protoplanetary disc. For a typical close-in planet at $0.1~\mathrm{AU}$, $M_{\mathrm{iso}}$ is about $0.6M_{\oplus}$ (Fig. \ref{fig:Fig1}). If a close-in planet forms \textit{in situ}, in order form a core of $M_{\mathrm{core}}\gtrsim M_{\oplus}$, a planetary embryo must undergo an additional stage of assembly---giant impacts---in order to achieve observed masses. Giant impacts between protoplanets, however, can lead to atmospheric mass-loss by the planetary embryos. A collision between an impactor and a planetary embryo generates a shock wave that travels through the interior of the target core. This shock wave subsequently initiates ground motion over the surface of the core, which in turn launches a shock into the atmosphere. This can lead to ejection of all or part of the atmosphere (Fig. \ref{fig:Fig3}). In Section \ref{sec:LocalLoss}, we thus calculate the local atmospheric mass-loss fraction (which we denote $\chi_{\mathrm{local}}$) as a function of local ground velocity, and use these results to calculate the global atmospheric loss fraction $\chi_{\mathrm{global}}$ (Section \ref{sec:GlobalLoss}). Previous work has focused on atmospheric loss in thin or plane-parallel atmospheres \citep{GendaAbeIcarus,SchlichtingSari2014}. Here we consider the regime applicable to close-in exoplanets, in which the envelope radius is a substantial fraction of the core radius and in which curvature effects become non-negligible. 

\subsection{Local Atmospheric Loss}\label{sec:LocalLoss}
In order to determine the mass-loss fraction of an envelope as a function of ground velocity and atmospheric mass, we perform numerical simulations of ground motion-induced shock propagation through an atmosphere. The Lagrangian equations of motion of the atmosphere are
\begin{subequations}\label{eq:HydroEqns}
\begin{align}
\frac{Dr}{Dm} 	&=  \frac{1}{4\pi r^2 \rho},\\
\frac{Du}{Dt}  	&= -4\pi r^2 \frac{\partial p}{\partial m} - 
					\frac{GM_{\mathrm{core}}}{r^2} - \\ \nonumber 
				&   4\pi G\int_{R_{\mathrm{core}}}^r \rho M(<r) dr,\\
\frac{Dr}{Dt}  	&= u, 
\end{align}
\end{subequations}
where $r$ is the distance from the centre of the core, $m$ is the mass enclosed in a mass shell at $r$, $\rho$ is the bulk density of the atmosphere, $u$ is the velocity, $p$ is pressure, and $t$ is time. Here, $M\left(<r\right)$ is the atmospheric mass interior to $r$. Since radiative losses are negligible, we assume that the shock propagates adiabatically, in which case the equation of state is
\begin{equation}\label{eq:HydroEqnsState}
\frac{Dp}{Dt}   = -4\pi \rho \gamma p \frac{\partial \left(r^2 u\right)}{\partial m},
\end{equation}
where $\gamma$ is the adiabatic index of the fluid flow. We work within a Lagrangian framework, so that parametrizing the hydrodynamic equations in terms of shell mass $m$ allows us to track each parcel and to determine if it is lost due to escape. The criterion for hydrodynamic escape is that the parcel at some $t \gg t_0$ has a velocity greater than its initial radius-dependent escape velocity at $t = t_0$, where $t_0$ is the time at which the shock is launched into the envelope. In order to integrate the equations of motion, we must specify the boundary condition at the base of the atmosphere. Here, following \citet{GendaAbeIcarus}, we assume that at $t = t_0$, the ground velocity is given by $v_g$, after which its time evolution is governed by $\partial v_g/\partial t = -GM_{\mathrm{core}}/r^2$.

In Section \ref{sec:IsoMassResults}, we showed that in the inner-disc regime applicable to close-in planets ($\lesssim 0.1~\mathrm{AU}$), the atmospheres can be regarded as adiabatic. For our atmospheric-loss calculations, we therefore assume an adiabatic profile [Eqs. \eqref{eq:adiabatic_eqs}] as the initial condition for Eqs. \eqref{eq:HydroEqns} and \eqref{eq:HydroEqnsState}. The adiabatic index $\gamma_a$ of Eqs. \eqref{eq:adiabatic_eqs} is not necessarily the same as $\gamma$ in the hydrodynamic equations. In particular, $\gamma$ depends on the ionization state of the gas. In our hydrodynamic simulations, we find that the dependence of the solutions on $\gamma$ is relatively weak, consistent with results from the literature \citep{GendaAbeIcarus,SchlichtingSari2014}. We thus take $\gamma = \gamma_a = 7/5$. If we assume an initial adiabatic profile for the atmosphere, subsequent nondimensionalization of the equations of motion shows that the solutions can be parametrized in terms of the envelope to core mass ratio $M_{\mathrm{atm}}/M_{\mathrm{core}}$ and the ratio of the ground velocity to the surface escape velocity $v_g/v_{\mathrm{esc}}$.  

\begin{figure}
	\centering
		\includegraphics[width=0.5\textwidth]{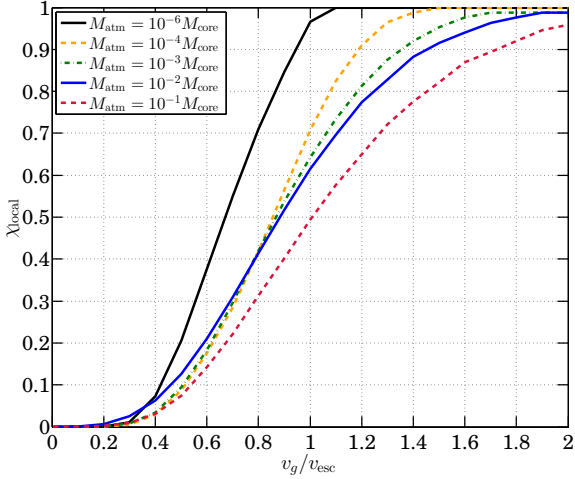}
	\caption{
		Local atmospheric mass-loss fraction as a function of ground 
	velocity for different envelope to core mass ratios. We show five cases 
	corresponding to different atmosphere-to-core mass ratios $M_{\mathrm{atm}}/M_{\mathrm{core}}$: 
	$M_{\mathrm{atm}} = 10^{-6}M_{\mathrm{core}}$, 
	$M_{\mathrm{atm}} = 10^{-4}M_{\mathrm{core}}$,
	$M_{\mathrm{atm}} = 10^{-3}M_{\mathrm{core}}$,
	$M_{\mathrm{atm}} = 10^{-2}M_{\mathrm{core}}$, and
	$M_{\mathrm{atm}} = 10^{-1}M_{\mathrm{core}}$.
	For the thinnest envelope, $M_{\mathrm{atm}} = 10^{-6}M_{\mathrm{core}}$, our results agree with the plane-parallel simulations of \citet{SchlichtingSari2014} and \citet{GendaAbeIcarus}.
	}
	\label{fig:Fig4}
\end{figure}

We solve Eqs. \eqref{eq:HydroEqns} and \eqref{eq:HydroEqnsState} numerically using a finite-difference, staggered grid scheme with artificial viscosity. The radial dependence is discretized over 300 mesh points, and an adaptive time step is implemented to ensure numerical stability. Integrations are carried out over several hundred thousand time steps in order to ensure convergence. In Fig. \ref{fig:Fig4}, we show the local atmospheric mass-loss fraction $\chi_{\mathrm{local}}$ as a function of $v_g/v_{\mathrm{esc}}$ for $M_{\mathrm{atm}}/M_{\mathrm{core}}$ ratios spanning $10^{-6}-10^{-1}$. 

\subsection{Global Atmospheric Loss}\label{sec:GlobalLoss}
To calculate the global mass-loss due to an impact, we must determine how the ground velocity varies with location on the planetary surface. Given our general ignorance of the internal structure of planetary embryos, we appeal to a simple analytical model which uses momentum conservation of an impact-induced shock as it propagates through the core.  Momentum conservation of the shock is consistent with results from smooth particle hydrodynamic simulations of catastrophic impacts \citep{LoveAhrens,BenzAsphaug}. In particular, suppose an impactor of mass $m_{\mathrm{imp}}$ collides with a protoplanet of mass $M_{\mathrm{core}}$ at a speed of $v_{\mathrm{imp}}$. Then if $Q^{*} \equiv m_{\mathrm{imp}}v_{\mathrm{imp}}^2/\left(2M_{\mathrm{core}}\right)$ is the specific energy of the impactor, a scaling consistent with constant impactor speed and momentum conservation of the shock yields $Q^{*} \propto R_{\mathrm{core}}$, close to the $Q^{*} \propto R_{\mathrm{core}}^{1.13}$ reported by \citet{LoveAhrens}. (A similar scaling assuming energy conservation of the shock yields $Q^{*} \propto R_{\mathrm{core}}^{2}$.)

\begin{figure}
	\centering
		\includegraphics[width=0.5\textwidth]{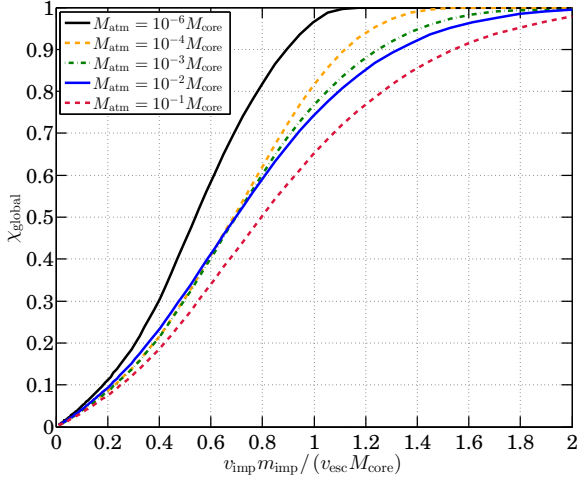}
	\caption{Global atmospheric mass-loss fraction as a function of 
	$v_{\mathrm{imp}}m_{\mathrm{imp}}/\left(v_{\mathrm{esc}}M_{\mathrm{core}}\right)$. 
	Each curve represents a solution to Eq. \eqref{eq:globalLossDef} for given 
	envelope mass to core mass ratio.}
	\label{fig:Fig5}
\end{figure}

If we consider the impact to be a point-like explosion, the shock wave will propagate radially from the point of impact. The total volume swept out by a shock that has traversed a distance $l$ from the point of impact is then $4\pi/3 \times R_{\mathrm{core}}^3 (l/2R_{\mathrm{core}})^3 [ 4 - 3(l/2R_{\mathrm{core}})]$. This volume is equivalent to that enclosed by the intersection of two spheres, one of radius $R_{\mathrm{core}}$ and the other (centred at the point of impact) of radius $l$ (Fig. \ref{fig:Fig3}). If this volume propagates with a speed of $v_s$, and if we assume that the core is of constant density, then setting the momentum of the swept-up mass equal to the impactor momentum $m_{\mathrm{imp}}v_{\mathrm{imp}}$ yields for the speed of the shocked material
\begin{equation}
v_s = v_{\mathrm{imp}}\left(\frac{m_{\mathrm{imp}}}{M_{\mathrm{core}}}\right)
			\left(\frac{l}{2R_{\mathrm{core}}}\right)^{-3}\left[4 - 3\left(\frac{l}{2R_{\mathrm{core}}}\right) \right]^{-1}.
\end{equation}
A constant density core is a reasonable first-order approximation given our general ignorance of the interior structure of these bodies during their formation and given the fact that the temperatures at the base of atmospheres have typical values of several thousands to $\sim 10,000~\mathrm{K}$. These temperatures are so high that the cores should be well mixed and not differentiated during the giant impact phase. The ground speed $v_g$, which is the projection of $v_s$ along the surface normal of the core for a given $l$, is then given by $v_g = v_s\left(l/2R_{\mathrm{core}}\right)$, so that after normalizing the ground speed against the escape velocity of the core, we get
\begin{equation}\label{eq:v_g_v_esc_phi}
\frac{v_g}{v_{\mathrm{esc}}} = \left(\frac{v_{\mathrm{imp}}}{v_{\mathrm{esc}}} \right) 
		\left(\frac{m_{\mathrm{imp}}}{M_{\mathrm{core}}}\right)
		\left(\frac{l}{2R_{\mathrm{core}}}\right)^{-2}\left[4 - 3\left(\frac{l}{2R_{\mathrm{core}}}\right) \right]^{-1}.
\end{equation}
The factor $l/2R_{\mathrm{core}}$ can be rewritten in terms of the angle $\phi$ subtended on the spherical surface from the point of impact to a given ground location through the relation $l/2R_{\mathrm{core}} = \sqrt{(1 - \cos \phi)/2}$ (Fig. \ref{fig:Fig3}). Since the atmospheric mass-loss fraction is parametrized in terms of the ratio of atmospheric mass to core mass $M_{\mathrm{atm}}/M_{\mathrm{core}}$ (Section \ref{sec:LocalLoss}), the global atmospheric mass-loss fraction can be written as
\begin{align}\label{eq:globalLossDef}
\chi_{\mathrm{global}}\left(M_{\mathrm{atm}}/M_{\mathrm{core}}\right)= 
\frac{1}{2} \int_{0}^{\pi} \chi_{\mathrm{local}}\left(v_g/v_{\mathrm{esc}},
						M_{\mathrm{atm}}/M_{\mathrm{core}},\phi\right)\times  \nonumber \\
						\sin\phi d\phi.
\end{align}
Combining the results in Fig. \ref{fig:Fig4} with Eq. \eqref{eq:v_g_v_esc_phi}, we perform the integration given in Eq. \eqref{eq:globalLossDef} to determine the global mass-loss as a function of $v_{\mathrm{imp}}m_{\mathrm{imp}}/\left(v_{\mathrm{esc}}M_{\mathrm{core}}\right)$ (Fig. \ref{fig:Fig5}). 

\section{Atmospheric Erosion Through the Giant Impact Phase}\label{sec:CollModel}

\begin{figure}
	\centering
		\includegraphics[width=0.5\textwidth]{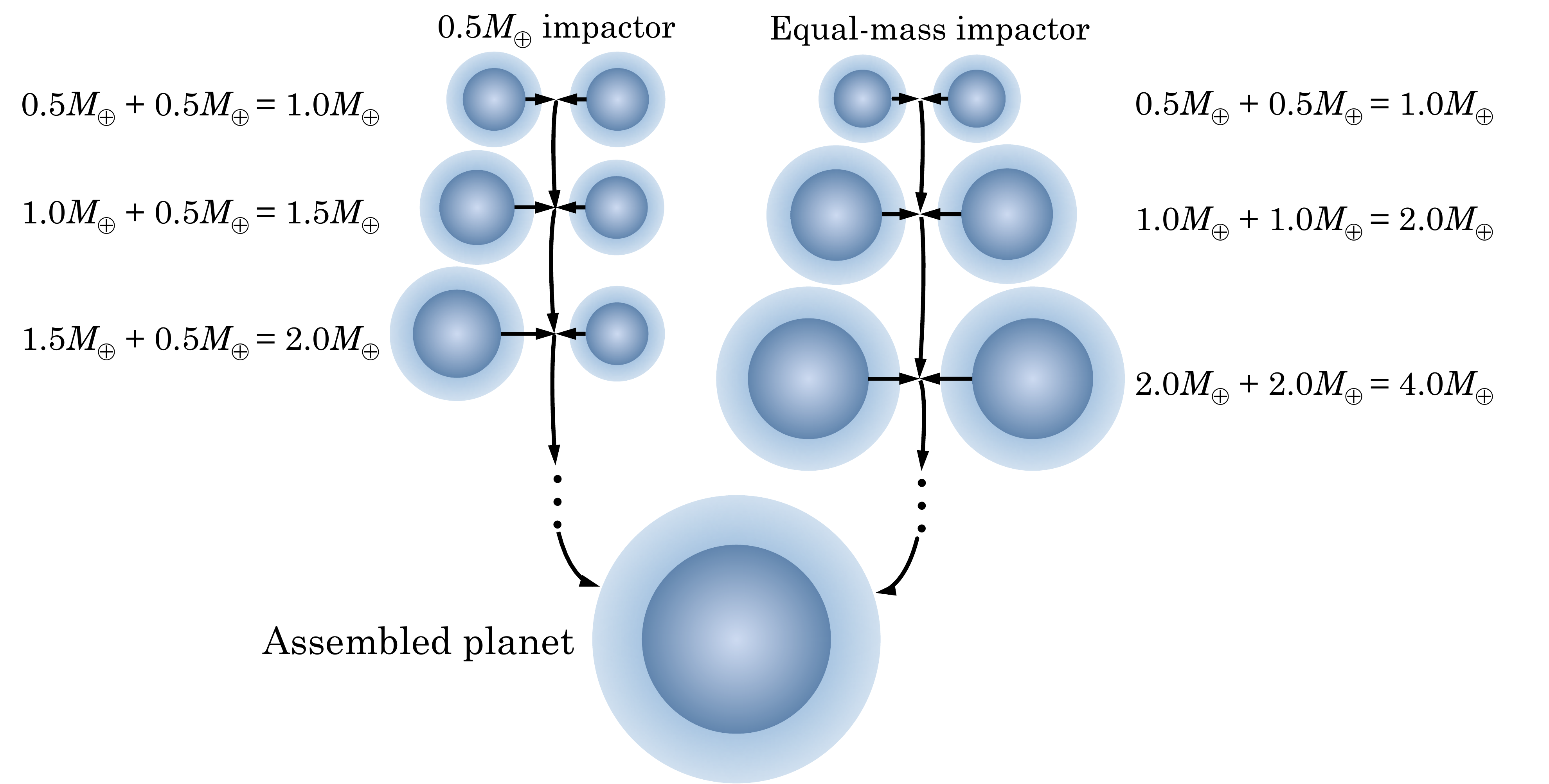}
	\caption{Example impact histories. On the left-hand side, each impactor has a mass 
	of $0.5M_{\oplus}$, and the planet core grows gradually by successive addition 
	of $0.5M_{\oplus}$ impactors. On the right-hand side, each impact consists of a 
	collision between equal-mass impactors. Each giant impact history represents 
	an extreme formation scenario, with a real giant impact history	likely to be 
	some combination of the two.}
	\label{fig:Fig6}
\end{figure}

To study the atmospheric mass-loss during the giant impact phase, we construct hypothetical collision histories using the results from Section \ref{sec:HydroEscape}. Fig. \ref{fig:Fig6} shows schematically two impact scenarios that we investigate. On the left side, a series of successive impacts with $0.5M_{\oplus}$ embryos takes place, while on the right, a series of impacts occurs between equal-mass impactors. The impact histories that we investigate here represent two extremes: one in which the impactor mass is constant such that $M_{\mathrm{core}}/m_{\mathrm{imp}}$ always increases, and one in which equal-size impacts occur ($M_{\mathrm{core}}/m_{\mathrm{imp}} = 1$). The actual impact history is likely to be a combination of these two scenarios. We determine the evolution of the atmosphere-to-core mass ratio $M_{\mathrm{atm}}/M_{\mathrm{core}}$ with successive impacts. For impacts between disparate mass bodies, we assume a relative random velocity $v_{\mathrm{ran}} \sim v_{\mathrm{esc}}$, so that the impact velocity $v_{\mathrm{imp}}\sim \sqrt{2} v_{\mathrm{esc}}$. To calculate the global atmospheric mass-loss of the target, we use the results from Section \ref{sec:GlobalLoss} to determine the atmospheric mass-loss fraction for the larger body. For the smaller impactor, we assume that all its atmospheric mass is lost. For impacts between equal-sized bodies, there is some ambiguity in how the impact and subsequent atmospheric loss occurs. In this case, we assume that impacts occur at a relative velocity of  $v_{\mathrm{imp}}\sim \sqrt{2} v_{\mathrm{esc}}$, where $v_{esc}$ is the escape velocity of a single body, and that a shock is launched into each body with a velocity about half the impact velocity, such that $v_{\mathrm{imp}}m_{\mathrm{imp}}/\left(v_{\mathrm{esc}}M_{\mathrm{core}}\right) \approx 0.7$. We assume that the resulting planet has an envelope with a mass equal to the sum of the remaining envelopes of the two individual impactors.

\begin{figure}
	\centering
		\includegraphics[width=0.5\textwidth]{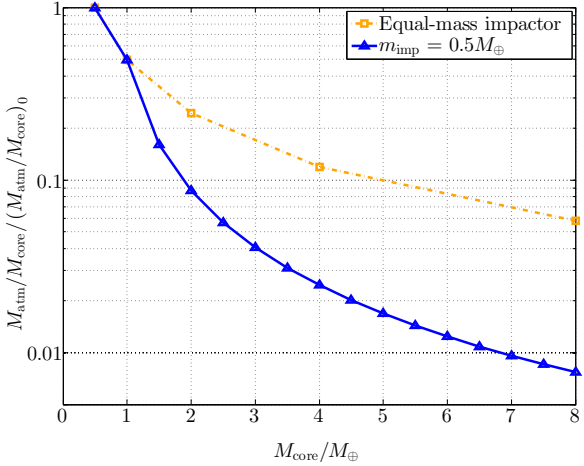}
	\caption{
	Evolution of atmosphere-to-core mass ratio as a function of impact 
	history. Each marker indicates a different collision event. Blue triangles 
	indicate a collision history in which the embryo grows gradually through 
	impacts with $0.5M_{\oplus}$ impactors (left-hand side of Fig. \ref{fig:Fig6}). 
	The orange squares are the case in which all impacts occur between	equal-mass bodies (right 
	side of Fig. \ref{fig:Fig6}). Atmosphere-to-core mass ratios are 
	normalized by the initial isolation mass atmosphere-to-core mass ratio, $\left(M_{\mathrm{atm}}/M_{\mathrm{core}}\right)_0$, which are 
	calculated in Section \ref{sec:IsoAccrete}. If there is no atmospheric
	loss throughout the giant impact history, then $M_{\mathrm{atm}}/M_{\mathrm{core}}/\left(M_{\mathrm{atm}}/M{_\mathrm{core}}\right)_0 = 1$.
	}
	\label{fig:Fig7}
\end{figure}

In Fig. \ref{fig:Fig7}, we show an example of the evolution of $M_{\mathrm{atm}}/M_{\mathrm{core}}$ for each of the two impact scenarios. In this example, we assume that the initial core mass is $0.5M_{\oplus}$ with the initial atmosphere-to-core mass ratio $\left(M_{\mathrm{atm}}/M_{\mathrm{core}}\right)_0 = 2 \times 10^{-3}$ (see Fig. \ref{fig:Fig2}). Each marker indicates a separate impact event. The blue triangles indicate the history in which $M_{\mathrm{core}}/m_{\mathrm{imp}}$ always increases. The orange squares represent the case in which all collisions are between equal-mass impactors. If no atmosphere is lost throughout the giant impact history, $M_{\mathrm{atm}}/M_{\mathrm{core}}$ equals $\left(M_{\mathrm{atm}}/M_{\mathrm{core}}\right)_0$. For a final planet mass of $4.5M_{\oplus}$ and a collision history in which all impactors are $0.5M_{\oplus}$, our model yields a final atmosphere-to-core mass ratio of $\sim 10^{-2}\left(M_{\mathrm{atm}}/M_{\mathrm{core}}\right)_0 \sim 10^{-5}$. The case in which giant impacts occur between equal-mass impactors, on the other hand, yields a final $M_{\mathrm{atm}}/M_{\mathrm{core}} \sim 10^{-1}\left(M_{\mathrm{atm}}/M_{\mathrm{core}}\right)_0 \sim 10^{-4}$.

\begin{figure}
	\centering
		\includegraphics[width=0.5\textwidth]{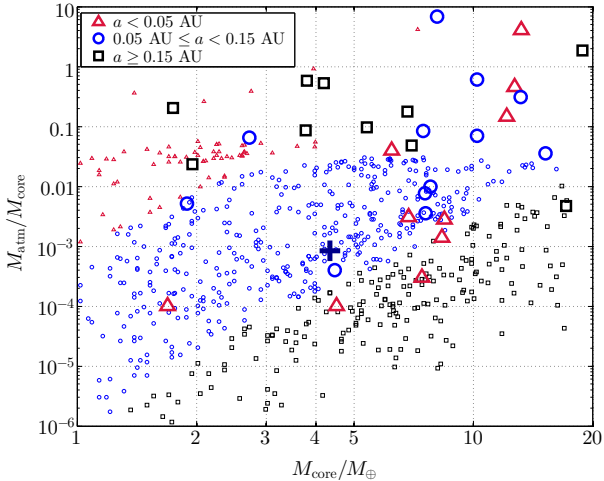}
	\caption{ Atmosphere-to-core mass ratios for observed close-in
	exoplanets after a phase of giant impacts. Modelled planets are 
	shown in the figure with small markers, while observed close-in planet 
	atmospheric masses and corresponding semimajor axis ranges \citep{LopezFortney} 
	are shown with large markers. Initial atmospheric masses are calculated 
	assuming observed close-in planets initially accreted gas envelopes 
	\textit{in situ} as isolation masses (see, e.g., Fig. \ref{fig:Fig1}) and 
	were then assembled by giant impacts. The giant impact results shown in this 
	figure are the mean of 10 simulations for each planet. The median atmosphere 
	to core mass ratio after giant impacts for a close-in planet is $8\times 10^{-4}$
	(blue cross), with a lower and upper quartile range of $1\times 10^{-4}$ and $6\times 10^{-3}$, 
	respectively.}
	\label{fig:Fig8}
\end{figure}

We have applied the collision histories shown in Fig. \ref{fig:Fig6} to observed exoplanets. In Fig. \ref{fig:Fig8}, we show the atmospheric masses that we obtain for observed close-in exoplanets after a phase of giant impacts. For each \textit{Kepler} planet, we take its calculated isolation mass (see, e.g., Fig. \ref{fig:Fig1}) and calculate the initial atmospheric mass accreted by the isolation mass core given its semimajor axis and the corresponding enhancement in gas density relative to the MMSN (see Fig. \ref{fig:Fig1}). We then perform Monte Carlo simulations to determine the atmospheric mass-loss due to giant impacts. For each planet, we conduct 10 trials. In each trial, a core of initial mass $M_{\mathrm{iso}}$ successively undergoes a series of giant impacts with impactors either of its own mass ($M_{\mathrm{core}}/m_{\mathrm{imp}} = 1$) or with a mass equal to the initial isolation mass $M_{\mathrm{iso}}$ ($M_{\mathrm{core}}/m_{\mathrm{imp}}$ increasing). For each collision, we randomly choose which type of impact occurs and assign a probability of $(M_{\mathrm{core}}/M_{\mathrm{iso}})/( M_{\mathrm{core}}/M_{\mathrm{iso}}+1)$ that an impact with an embryo of mass $M_{\mathrm{iso}}$ occurs. This probability is chosen such that on average, a planet grows equally in mass by merging with equal-size bodies and smaller ones. For each trial, the core undergoes giant impacts until the final, observed mass is assembled. The post-giant impact atmosphere-to-core mass ratios shown in Fig. \ref{fig:Fig8} are the mean of all 10 trials for each planet. The median atmosphere-to-core mass ratio after the phase of giant impacts for observed close-in planets is $8\times 10^{-4}$ and the values for the lower and upper quartile range are $1\times 10^{-4}$ and $6\times 10^{-3}$, respectively. These values are consistent with terrestrial planet atmospheres and exoplanets that have inferred rocky compositions, but are typically smaller, by an order of magnitude, than atmospheric masses of $1 - 10$ per cent inferred from observation for many close-in exoplanets. \textit{In situ} formation of close-in planets via giant impacts typically does not result in atmospheric masses that are 1--10 per cent or more of the core mass.

\section{Post-Giant Impact Accretion}\label{sec:PostGiant}
Since atmospheric masses containing 1--10 per cent or more of the total planet mass are difficult to achieve as a result of giant impacts, we investigate now the importance of gas accretion after assembly has taken place. In this case, the analysis presented in Section \ref{sec:IsoAccrete} still holds with slight modification. In order for giant impacts to proceed, the gas surface density in the disc will have had to decrease, so that $\Sigma_{g} \sim \Sigma_{s}$ \citep{Goldreich2004}. In the full disc, the excess of gas relative to solids can effectively limit giant impacts from proceeding by damping out the large eccentricities required for them. It is only after a sufficient amount of gas has been dissipated from the disc that giant impacts can proceed. Therefore, in order to calculate the masses of envelopes accreted after giant impacts, we take the gas surface density to be 200 times smaller than before the giant impact phase. For our opacity calculations, we assume that the gas-to-dust ratio is still 200.

\begin{figure}
	\centering
		\includegraphics[width=0.5\textwidth, angle = 0]{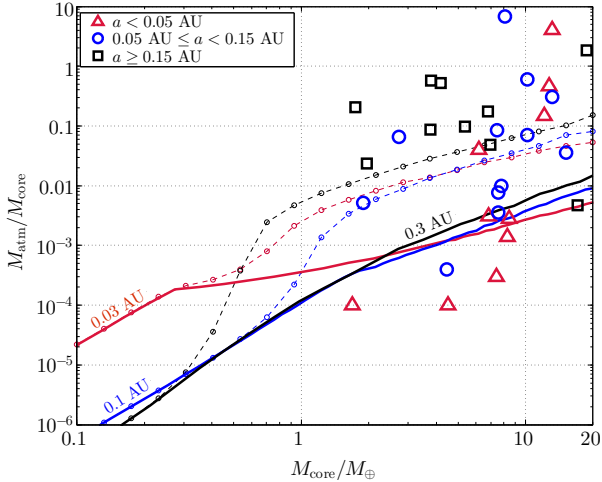}
	\caption{Atmospheric masses due to gas envelope accretion by assembled core masses 
after giant impacts. $M_{\mathrm{atm}}/M_{\mathrm{core}}$ for observed 
close-in planets are shown with triangles (semimajor axes less than $0.05~\mathrm{AU}$), 
circles ($0.05-0.15~\mathrm{AU}$), and squares (semimajor axes greater than 
$0.15~\mathrm{AU}$)\citep{LopezFortney}. $M_{\mathrm{atm}}/M_{\mathrm{core}}$ 
ratios calculated for a range of core masses are shown at semimajor axes of 
$0.03~\mathrm{AU}$ (red lines), $0.1~\mathrm{AU}$ (blue lines), and $0.3~\mathrm{AU}$ (black
lines) for two limiting cases. The thick solid lines correspond to the case in which $L = L_{\mathrm{acc}}$, where the accretion luminosity is due to the gravitational potential energy from a giant impact between two equal mass bodies of $0.5 M_{\mathrm{core}}$ that is released over $0.8~\mathrm{Myr}$. The thin dashed lines with circular markers corresponds to the case in which $L_{\mathrm{acc}} = 0$.}
	\label{fig:Fig9}
\end{figure}

In Fig. \ref{fig:Fig9}, we show $M_{\mathrm{atm}}/M_{\mathrm{core}}$ for gas accretion after giant impacts for two limiting cases. In thick solid lines, we show the case in which $L = L_{\mathrm{acc}}$. Here, we assume that the gravitational potential energy resulting from the last mass doubling of the planet by giant impacts is released over the disc dissipation time-scale, so that $\dot{M}_{\mathrm{core}} = 0.5M_{\mathrm{core}}/\tau_{\mathrm{diss}}$, where we take the disc dissipation time-scale $\tau_{\mathrm{diss}} \sim 800~\mathrm{kyr}$ \citep{Hillenbrand}. In thin dashed lines with circular markers, we show the case in which $L_{\mathrm{acc}} = 0$ and the evolution of $L$ is governed by Eq. \eqref{eq:entevo}. In this case, we cut off gas accretion at $\tau_{\mathrm{diss}}$.

In contrast to envelope accretion from a full gas disc investigated in Section \ref{sec:IsoAccrete}, in the case of a dissipating gas disc, the gas densities and opacities are now sufficiently low such that the envelope can cool, contract, and accrete more atmospheric mass over the disc dissipation time-scale if $L_{\mathrm{acc}}=0$. In this case, atmospheres containing several per cent of the planets' total mass can be accreted. The $L_{\mathrm{acc}}=0$ case is likely an upper limit since planetesimal accretion very likely continued after the giant impact stage and since the giant impacts themselves will give rise to significant core luminosity. The magnitude of the core luminosity is highly uncertain since it depends on the viscosity of the core, which is unknown. If the gravitational potential energy resulting from the last mass doubling of the planet is released over the disc dissipation time-scale, then the accreted envelope masses are reduced by about an order of magnitude compared to the $L_{\mathrm{acc}}=0$ case (see Fig. \ref{fig:Fig9}).

While the two limiting cases we explore here show a plausible range of atmospheric masses, atmospheric masses exceeding several per cent seem difficult to accrete from a reduced gas disc after giant impacts. Post-giant impact accretion does not seem to be capable of producing atmospheric masses exceeding several per cent of the core mass.

\section{Discussion and Conclusions}\label{sec:DiscConc}

How and where close-in super-Earths and mini-Neptunes formed is one of the outstanding problems in planet formation. It has been proposed that close-in super-Earths and mini-Neptunes formed \textit{in situ} either by delivery of $50-100M_{\oplus}$ of rocky material into the inner disc \citep{HansenMurray}, or in a disc enhanced relative to the MMSN \citep{ChiangLaughlin}. In both cases, the final assembly of the planets occurs via giant impacts.

Here, we investigated the atmospheric masses that close-in planets can achieve before, maintain during, and accrete after giant impacts and compared these atmospheric masses with observations. First, we have found that the atmospheres accreted by isolation masses are small. For a typical close-in planet isolation mass of $\sim 0.6M_{\oplus}$ at 0.1~AU (Fig. \ref{fig:Fig1}), the atmosphere-to-core mass ratio $M_{\mathrm{atm}}/M_{\mathrm{core}}$ is about $10^{-3}-10^{-2}$ (Fig. \ref{fig:Fig2}). This value is already less than atmospheric masses of up to 1--10 per cent inferred for a significant fraction of close-in super-Earths and mini-Neptunes. Additionally, the $M_{\mathrm{atm}}/M_{\mathrm{core}}$ ratio is reduced further by a factor of $\sim 10^{-2}-10^{-1}$ due to giant impacts, leading to median atmosphere-to-core mass ratio after giant impacts of $8\times 10^{-4}$ with a lower and upper quartile range of $1\times 10^{-4}$ and $6\times 10^{-3}$, respectively. Such values are consistent with terrestrial planet atmospheres but more than an order of magnitude below atmospheric masses of $1-10$ per cent of the total planet mass inferred for many close-in exoplanets. Finally, we considered the accretion of gas envelopes by fully-assembled cores after the phase of giant impacts. In this case, we have found that in the best case scenario where there is no core luminosity from giant impacts, accreted atmospheric masses are at best several per cent (Fig. \ref{fig:Fig9}). If the gravitational potential energy resulting from the last mass doubling of the planet by giant impacts is released over the disc dissipation time-scale as core luminosity, then the accreted envelope masses are reduced by about an order of magnitude compared to the $L_{\mathrm{acc}}=0$ case. We note that the atmospheric masses we have calculated here should be regarded as upper limits since we have ignored other mass-loss mechanisms such as photoevaporation \citep{LopezFortney_PhotoEvap1,SanzForcada} and tides \citep{GuMassLoss}.

\begin{figure}
	\centering
		\includegraphics[width=0.5\textwidth, angle = 0]{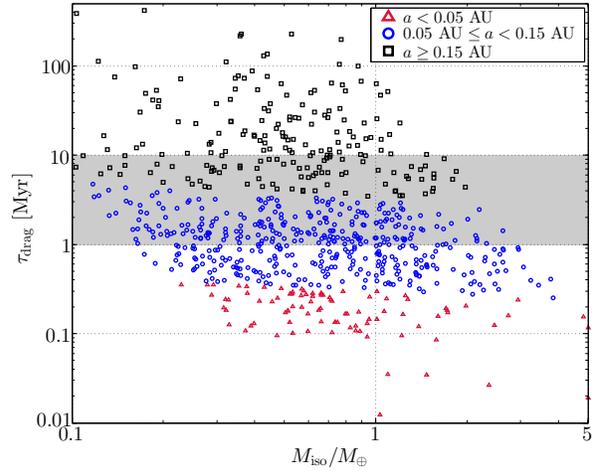}
	\caption{Radial drift time-scales $\tau_{\mathrm{drag}}$ due to gas drag for isolation masses calculated from close-in \textit{Kepler} planets (see Fig. \ref{fig:Fig1}). Markers are coloured based on observed semimajor axis. For many systems, the radial drift time-scale is short compared to typical disc time-scales of $1-10~\mathrm{Myr}$ (shaded region), suggesting that these isolation masses will drift into their host star before giant impacts proceed.}
	\label{fig:Fig10}
\end{figure}

In addition to the results presented above, another challenge with \textit{in situ} formation is the time-scale in the inner disc associated with radial drift of the isolation masses into their host stars due to gas drag. The time-scale for radial drift $\tau_{\mathrm{drag}}$ due to gas drag in the Stokes regime is
\begin{equation}
\tau_{\mathrm{drag}} = \frac{16}{\pi C_D n^2} \left(\frac{M_p}{R_p^2 \Sigma_g}\right)\left(\frac{a\Omega}{c_s}\right)^3 \frac{1}{\Omega},
\end{equation}
where $C_D \sim 1$ is the drag coefficient. Here, $n$ is the power law index that describes the radial pressure profile in the disc, such that $P_{\mathrm{disc}} \propto a^{-n}$. Using the disc model we detail in Section \ref{sec:ModelDetails}, we find that $n = 3$ when $a \leq 0.1~\mathrm{AU}$ and that $n = 10/3$ when $a > 0.1~\mathrm{AU}$. In Fig. \ref{fig:Fig10}, we show $\tau_{\mathrm{drag}}$ for observed close-in exoplanet isolation masses at their observed semimajor axes. To calculate $\tau_{\mathrm{drag}}$, we use the isolation masses and density enhancements relative to the MMSN shown in Fig. \ref{fig:Fig1}, while to calculate the radius of the isolation mass, we use Eq. \eqref{eq:MassRadius}. Fig. \ref{fig:Fig10} demonstrates that a large fraction of isolation masses is expected to drift into their host stars for typical disc lifetimes ranging from $1-10~\mathrm{Myr}$ (shaded region). Thus, even without accounting for type I migration, this short time-scale for inward radial drift due to gas drag presents an impediment to \textit{in situ} formation.

Lastly, while we have found that it is challenging to form close-in planets with atmospheric masses that are at least several per cent of the their total mass \textit{in situ} with giant impacts, it may be possible that such planets reached their fully-assembled cores by planetesimal accretion facilitated by radial drift, thereby bypassing the giant impact phase altogether \citep{Lambrechts}. Recent work, however, suggests that planetesimal accretion results in oligarchic-type growth similar to that which also leads to the formation of isolation masses \citep{KretkeLevison}. 

Given the challenges with \textit{in situ} formation of close-in planets with massive atmospheres discussed above, we favour the possibility that they formed further out in the disc and migrated inwards to their current location. Formation at semimajor axes of a few AU does not require significantly enhanced disc masses compared to the MMSN \citep{Schlichting2014} and the formation time-scale, migration time-scale and disc lifetime are comparable, circumventing the need for extreme fine tuning.

\section*{Acknowledgements}
We thank Re'em Sari for stimulating scientific discussions and the anonymous referee for valuable comments that helped to improve the manuscript.

\bibliographystyle{mn2e}
\bibliography{Atm_Loss}

\begin{thebibliography}{}
\makeatletter
\relax
\def\mn@urlcharsother{\let\do\@makeother \do\$\do\&\do\#\do\^\do\_\do\%\do\~}
\def\mn@doi{\begingroup\mn@urlcharsother \@ifnextchar [ {\mn@doi@}
  {\mn@doi@[]}}
\def\mn@doi@[#1]#2{\def\@tempa{#1}\ifx\@tempa\@empty \href
  {http://dx.doi.org/#2} {doi:#2}\else \href {http://dx.doi.org/#2} {#1}\fi
  \endgroup}
\def\mn@eprint#1#2{\mn@eprint@#1:#2::\@nil}
\def\mn@eprint@arXiv#1{\href {http://arxiv.org/abs/#1} {{\tt arXiv:#1}}}
\def\mn@eprint@dblp#1{\href {http://dblp.uni-trier.de/rec/bibtex/#1.xml}
  {dblp:#1}}
\def\mn@eprint@#1:#2:#3:#4\@nil{\def\@tempa {#1}\def\@tempb {#2}\def\@tempc
  {#3}\ifx \@tempc \@empty \let \@tempc \@tempb \let \@tempb \@tempa \fi \ifx
  \@tempb \@empty \def\@tempb {arXiv}\fi \@ifundefined
  {mn@eprint@\@tempb}{\@tempb:\@tempc}{\expandafter \expandafter \csname
  mn@eprint@\@tempb\endcsname \expandafter{\@tempc}}}

\bibitem[\protect\citeauthoryear{{Adams}, {Seager}  \& {Elkins-Tanton}}{{Adams}
  et~al.}{2008}]{AdamsSeager}
{Adams} E.~R.,  {Seager} S.,   {Elkins-Tanton} L.,  2008, \mn@doi [\apj]
  {10.1086/524925}, \href {http://adsabs.harvard.edu/abs/2008ApJ...673.1160A}
  {673, 1160}

\bibitem[\protect\citeauthoryear{{Agnor}, {Canup}  \& {Levison}}{{Agnor}
  et~al.}{1999}]{Agnor1999}
{Agnor} C.~B.,  {Canup} R.~M.,   {Levison} H.~F.,  1999, \mn@doi [\icarus]
  {10.1006/icar.1999.6201}, \href
  {http://adsabs.harvard.edu/abs/1999Icar..142..219A} {142, 219}

\bibitem[\protect\citeauthoryear{{Batalha} et~al.,}{{Batalha}
  et~al.}{2013}]{Batalha2013}
{Batalha} N.~M.,  et~al., 2013, \mn@doi [\apjs] {10.1088/0067-0049/204/2/24},
  \href {http://adsabs.harvard.edu/abs/2013ApJS..204...24B} {204, 24}

\bibitem[\protect\citeauthoryear{{Benz} \& {Asphaug}}{{Benz} \&
  {Asphaug}}{1999}]{BenzAsphaug}
{Benz} W.,  {Asphaug} E.,  1999, \mn@doi [\icarus] {10.1006/icar.1999.6204},
  \href {http://adsabs.harvard.edu/abs/1999Icar..142....5B} {142, 5}

\bibitem[\protect\citeauthoryear{{Bodenheimer} \& {Lissauer}}{{Bodenheimer} \&
  {Lissauer}}{2014}]{BodenheimerLissauer}
{Bodenheimer} P.,  {Lissauer} J.~J.,  2014, \mn@doi [\apj]
  {10.1088/0004-637X/791/2/103}, \href
  {http://adsabs.harvard.edu/abs/2014ApJ...791..103B} {791, 103}

\bibitem[\protect\citeauthoryear{{Borucki} et~al.,}{{Borucki}
  et~al.}{2011}]{Borucki2011}
{Borucki} W.~J.,  et~al., 2011, \mn@doi [\apj] {10.1088/0004-637X/736/1/19},
  \href {http://adsabs.harvard.edu/abs/2011ApJ...736...19B} {736, 19}

\bibitem[\protect\citeauthoryear{{Chambers}}{{Chambers}}{2001}]{Chambers2001}
{Chambers} J.~E.,  2001, \mn@doi [\icarus] {10.1006/icar.2001.6639}, \href
  {http://adsabs.harvard.edu/abs/2001Icar..152..205C} {152, 205}

\bibitem[\protect\citeauthoryear{{Chambers} \& {Wetherill}}{{Chambers} \&
  {Wetherill}}{1998}]{ChambersWetherill1998}
{Chambers} J.~E.,  {Wetherill} G.~W.,  1998, \mn@doi [\icarus]
  {10.1006/icar.1998.6007}, \href
  {http://adsabs.harvard.edu/abs/1998Icar..136..304C} {136, 304}

\bibitem[\protect\citeauthoryear{{Chiang} \& {Laughlin}}{{Chiang} \&
  {Laughlin}}{2013}]{ChiangLaughlin}
{Chiang} E.,  {Laughlin} G.,  2013, \mn@doi [\mnras] {10.1093/mnras/stt424},
  \href {http://adsabs.harvard.edu/abs/2013MNRAS.431.3444C} {431, 3444}

\bibitem[\protect\citeauthoryear{{D'Alessio}, {Calvet}  \&
  {Hartmann}}{{D'Alessio} et~al.}{2001}]{DAlessio}
{D'Alessio} P.,  {Calvet} N.,   {Hartmann} L.,  2001, \mn@doi [\apj]
  {10.1086/320655}, \href {http://adsabs.harvard.edu/abs/2001ApJ...553..321D}
  {553, 321}

\bibitem[\protect\citeauthoryear{{Genda} \& {Abe}}{{Genda} \&
  {Abe}}{2003}]{GendaAbeIcarus}
{Genda} H.,  {Abe} Y.,  2003, \mn@doi [\icarus]
  {10.1016/S0019-1035(03)00101-5}, \href
  {http://adsabs.harvard.edu/abs/2003Icar..164..149G} {164, 149}

\bibitem[\protect\citeauthoryear{{Goldreich}, {Lithwick}  \&
  {Sari}}{{Goldreich} et~al.}{2004}]{Goldreich2004}
{Goldreich} P.,  {Lithwick} Y.,   {Sari} R.,  2004, \mn@doi [\araa]
  {10.1146/annurev.astro.42.053102.134004}, \href
  {http://adsabs.harvard.edu/abs/2004ARA%26A..42..549G} {42, 549}

\bibitem[\protect\citeauthoryear{{Gu}, {Lin}  \& {Bodenheimer}}{{Gu}
  et~al.}{2003}]{GuMassLoss}
{Gu} P.-G.,  {Lin} D.~N.~C.,   {Bodenheimer} P.~H.,  2003, \mn@doi [\apj]
  {10.1086/373920}, \href {http://adsabs.harvard.edu/abs/2003ApJ...588..509G}
  {588, 509}

\bibitem[\protect\citeauthoryear{{Hansen} \& {Murray}}{{Hansen} \&
  {Murray}}{2012}]{HansenMurray}
{Hansen} B.~M.~S.,  {Murray} N.,  2012, \mn@doi [\apj]
  {10.1088/0004-637X/751/2/158}, \href
  {http://adsabs.harvard.edu/abs/2012ApJ...751..158H} {751, 158}

\bibitem[\protect\citeauthoryear{{Hayashi}}{{Hayashi}}{1981}]{Hayashi}
{Hayashi} C.,  1981, \mn@doi [Progress of Theoretical Physics Supplement]
  {10.1143/PTPS.70.35}, \href
  {http://adsabs.harvard.edu/abs/1981PThPS..70...35H} {70, 35}

\bibitem[\protect\citeauthoryear{{Hillenbrand}}{{Hillenbrand}}{2005}]{Hillenbrand}
{Hillenbrand} L.~A.,  2005, in {Livio} M.,  ed.,  STScI Symposium Series Vol.
  19, A Decade of Discovery: Planets around other stars.

\bibitem[\protect\citeauthoryear{{Howard} et~al.,}{{Howard}
  et~al.}{2010}]{HowardMarcyJohnson}
{Howard} A.~W.,  et~al., 2010, \mn@doi [Science] {10.1126/science.1194854},
  \href {http://adsabs.harvard.edu/abs/2010Sci...330..653H} {330, 653}

\bibitem[\protect\citeauthoryear{{Ikoma} \& {Hori}}{{Ikoma} \&
  {Hori}}{2012}]{IkomaHori}
{Ikoma} M.,  {Hori} Y.,  2012, \mn@doi [\apj] {10.1088/0004-637X/753/1/66},
  \href {http://adsabs.harvard.edu/abs/2012ApJ...753...66I} {753, 66}

\bibitem[\protect\citeauthoryear{{Kretke} \& {Levison}}{{Kretke} \&
  {Levison}}{2014}]{KretkeLevison}
{Kretke} K.~A.,  {Levison} H.~F.,  2014, preprint, \href
  {http://adsabs.harvard.edu/abs/2014arXiv1409.4430K} {} (\mn@eprint {arXiv}
  {1409.4430})

\bibitem[\protect\citeauthoryear{{Lambrechts} \& {Johansen}}{{Lambrechts} \&
  {Johansen}}{2012}]{Lambrechts}
{Lambrechts} M.,  {Johansen} A.,  2012, \mn@doi [\aap]
  {10.1051/0004-6361/201219127}, \href
  {http://adsabs.harvard.edu/abs/2012A%26A...544A..32L} {544, A32}

\bibitem[\protect\citeauthoryear{{Lee}, {Chiang}  \& {Ormel}}{{Lee}
  et~al.}{2014}]{LeeChiangOrmel}
{Lee} E.~J.,  {Chiang} E.,   {Ormel} C.~W.,  2014, preprint, \href
  {http://adsabs.harvard.edu/abs/2014arXiv1409.3578L} {} (\mn@eprint {arXiv}
  {1409.3578})

\bibitem[\protect\citeauthoryear{{Lissauer} et~al.,}{{Lissauer}
  et~al.}{2011a}]{LissauerArch}
{Lissauer} J.~J.,  et~al., 2011a, \mn@doi [\apjs] {10.1088/0067-0049/197/1/8},
  \href {http://adsabs.harvard.edu/abs/2011ApJS..197....8L} {197, 8}

\bibitem[\protect\citeauthoryear{{Lissauer} et~al.,}{{Lissauer}
  et~al.}{2011b}]{LissauerPacked}
{Lissauer} J.~J.,  et~al., 2011b, \mn@doi [Nature] {10.1038/nature09760}, \href
  {http://adsabs.harvard.edu/abs/2011Natur.470...53L} {470, 53}

\bibitem[\protect\citeauthoryear{{Lopez} \& {Fortney}}{{Lopez} \&
  {Fortney}}{2013}]{LopezFortney_PhotoEvap1}
{Lopez} E.,  {Fortney} J.~J.,  2013, in American Astronomical Society Meeting
  Abstracts \#221. p. \#333.04

\bibitem[\protect\citeauthoryear{{Lopez} \& {Fortney}}{{Lopez} \&
  {Fortney}}{2014}]{LopezFortney}
{Lopez} E.~D.,  {Fortney} J.~J.,  2014, \mn@doi [\apj]
  {10.1088/0004-637X/792/1/1}, \href
  {http://adsabs.harvard.edu/abs/2014ApJ...792....1L} {792, 1}

\bibitem[\protect\citeauthoryear{{Lopez}, {Fortney}  \& {Miller}}{{Lopez}
  et~al.}{2012}]{LopezFortneyMiller}
{Lopez} E.~D.,  {Fortney} J.~J.,   {Miller} N.,  2012, \mn@doi [\apj]
  {10.1088/0004-637X/761/1/59}, \href
  {http://adsabs.harvard.edu/abs/2012ApJ...761...59L} {761, 59}

\bibitem[\protect\citeauthoryear{{Love} \& {Ahrens}}{{Love} \&
  {Ahrens}}{1996}]{LoveAhrens}
{Love} S.~G.,  {Ahrens} T.~J.,  1996, \mn@doi [\icarus]
  {10.1006/icar.1996.0195}, \href
  {http://adsabs.harvard.edu/abs/1996Icar..124..141L} {124, 141}

\bibitem[\protect\citeauthoryear{{Movshovitz}, {Bodenheimer}, {Podolak}  \&
  {Lissauer}}{{Movshovitz} et~al.}{2010}]{Movshovitz}
{Movshovitz} N.,  {Bodenheimer} P.,  {Podolak} M.,   {Lissauer} J.~J.,  2010,
  \mn@doi [\icarus] {10.1016/j.icarus.2010.06.009}, \href
  {http://adsabs.harvard.edu/abs/2010Icar..209..616M} {209, 616}

\bibitem[\protect\citeauthoryear{{Piso} \& {Youdin}}{{Piso} \&
  {Youdin}}{2014}]{PisoYoudin}
{Piso} A.-M.~A.,  {Youdin} A.~N.,  2014, \mn@doi [\apj]
  {10.1088/0004-637X/786/1/21}, \href
  {http://adsabs.harvard.edu/abs/2014ApJ...786...21P} {786, 21}

\bibitem[\protect\citeauthoryear{{Rafikov}}{{Rafikov}}{2006}]{Rafikov2006}
{Rafikov} R.~R.,  2006, \mn@doi [\apj] {10.1086/505695}, \href
  {http://adsabs.harvard.edu/abs/2006ApJ...648..666R} {648, 666}

\bibitem[\protect\citeauthoryear{{Raymond}, {Quinn}  \& {Lunine}}{{Raymond}
  et~al.}{2004}]{Raymond2004}
{Raymond} S.~N.,  {Quinn} T.,   {Lunine} J.~I.,  2004, \mn@doi [\icarus]
  {10.1016/j.icarus.2003.11.019}, \href
  {http://adsabs.harvard.edu/abs/2004Icar..168....1R} {168, 1}

\bibitem[\protect\citeauthoryear{{Rogers}, {Bodenheimer}, {Lissauer}  \&
  {Seager}}{{Rogers} et~al.}{2011}]{Rogers2011}
{Rogers} L.~A.,  {Bodenheimer} P.,  {Lissauer} J.~J.,   {Seager} S.,  2011,
  \mn@doi [\apj] {10.1088/0004-637X/738/1/59}, \href
  {http://adsabs.harvard.edu/abs/2011ApJ...738...59R} {738, 59}

\bibitem[\protect\citeauthoryear{{Safronov}}{{Safronov}}{1972}]{Safronov}
{Safronov} V.~S.,  1972, {Evolution of the protoplanetary cloud and formation
  of the earth and planets.}.
Keter Publishing House.

\bibitem[\protect\citeauthoryear{{Sanz-Forcada}, {Micela}, {Ribas}, {Pollock},
  {Eiroa}, {Velasco}, {Solano}  \& {Garc{\'{\i}}a-{\'A}lvarez}}{{Sanz-Forcada}
  et~al.}{2011}]{SanzForcada}
{Sanz-Forcada} J.,  {Micela} G.,  {Ribas} I.,  {Pollock} A.~M.~T.,  {Eiroa} C.,
   {Velasco} A.,  {Solano} E.,   {Garc{\'{\i}}a-{\'A}lvarez} D.,  2011, \mn@doi
  [\aap] {10.1051/0004-6361/201116594}, \href
  {http://adsabs.harvard.edu/abs/2011A%26A...532A...6S} {532, A6}

\bibitem[\protect\citeauthoryear{{Schlichting}}{{Schlichting}}{2014}]{Schlichting2014}
{Schlichting} H.~E.,  2014, \mn@doi [\apjl] {10.1088/2041-8205/795/1/L15},
  \href {http://adsabs.harvard.edu/abs/2014ApJ...795L..15S} {795, L15}

\bibitem[\protect\citeauthoryear{{Schlichting}, {Sari}  \&
  {Yalinewich}}{{Schlichting} et~al.}{2015}]{SchlichtingSari2014}
{Schlichting} H.~E.,  {Sari} R.,   {Yalinewich} A.,  2015, \mn@doi [\icarus]
  {10.1016/j.icarus.2014.09.053}, \href
  {http://adsabs.harvard.edu/abs/2015Icar..247...81S} {247, 81}

\bibitem[\protect\citeauthoryear{{Schneider}, {Dedieu}, {Le Sidaner}, {Savalle}
   \& {Zolotukhin}}{{Schneider} et~al.}{2011}]{ExoEuCat}
{Schneider} J.,  {Dedieu} C.,  {Le Sidaner} P.,  {Savalle} R.,   {Zolotukhin}
  I.,  2011, \mn@doi [\aap] {10.1051/0004-6361/201116713}, \href
  {http://adsabs.harvard.edu/abs/2011A%26A...532A..79S} {532, A79}

\bibitem[\protect\citeauthoryear{{Seager}, {Kuchner}, {Hier-Majumder}  \&
  {Militzer}}{{Seager} et~al.}{2007}]{SeagerMassRadius}
{Seager} S.,  {Kuchner} M.,  {Hier-Majumder} C.~A.,   {Militzer} B.,  2007,
  \mn@doi [\apj] {10.1086/521346}, \href
  {http://adsabs.harvard.edu/abs/2007ApJ...669.1279S} {669, 1279}

\bibitem[\protect\citeauthoryear{{Zhu}, {Hartmann}  \& {Gammie}}{{Zhu}
  et~al.}{2009}]{Zhu}
{Zhu} Z.,  {Hartmann} L.,   {Gammie} C.,  2009, \mn@doi [\apj]
  {10.1088/0004-637X/694/2/1045}, \href
  {http://adsabs.harvard.edu/abs/2009ApJ...694.1045Z} {694, 1045}

\makeatother
\end{thebibliography}

\label{lastpage}

\end{document}